# Polarized Electrons Using the PWT RF Gun[*]

J. E. Clendenin[†], R. Kirby[†], Y. Luo[¶], D. Newsham[¶], D. Yu[¶]

[†]*Stanford Linear Accelerator Center, 2575 Sand Hill Rd., Menlo Park, CA 94025, USA*
[¶]*DULY Research Inc., Rancho Palos Verdes, CA 90275, USA*

**Abstract**

Future colliders that require low-emittance highly-polarized electron beams are the main motivation for developing a polarized rf gun. However there are both technical and physics issues in generating highly polarized electron beams using rf guns that remain to be resolved. The PWT design offers promising features that may facilitate solutions to technical problems such as field emission and poor vacuum. Physics issues such as emission time now seem to be satisfactorily resolved. Other issues, such as the effect of magnetic fields at the cathode—both those associated with the rf field and those imposed by schemes to produce flat beams—are still open questions. Potential solution of remaining problems will be discussed in the context of the PWT design.

*Presented at the Workshop on Polarized Electron Sources and Polarimeters (PESP 2002), September 4-6, 2002, Danvers, MA.*

[*]Work supported by Department of Energy contract DE–AC03–76SF00515.

# Polarized Electrons Using the PWT RF Gun[*]

J. E. Clendenin[†], R. Kirby[†], Y. Luo[¶], D. Newsham[¶], D. Yu[¶]

[†]*Stanford Linear Accelerator Center, 2575 Sand Hill Rd., Menlo Park, CA 94025, USA*
[¶]*DULY Research Inc., Rancho Palos Verdes, CA 90275, USA*

**Abstract**. Future colliders that require low-emittance highly-polarized electron beams are the main motivation for developing a polarized rf gun. However there are both technical and physics issues in generating highly polarized electron beams using rf guns that remain to be resolved. The PWT design offers promising features that may facilitate solutions to technical problems such as field emission and poor vacuum. Physics issues such as emission time now seem to be satisfactorily resolved. Other issues, such as the effect of magnetic fields at the cathode—both those associated with the rf field and those imposed by schemes to produce flat beams—are still open questions. Potential solution of remaining problems will be discussed in the context of the PWT design.

## INTRODUCTION

Colliders presently being designed require ~ 1 nC of charge per microbunch with an emittance of $10^{-8}$ m in the vertical plane for both the electron and positron beams. The electron beam is required to be polarized. To date all polarized electron beams for accelerators have been produced by dc-biased photocathode guns. For a bias of 100 kV, the typical transverse emittance for a solid-state photocathode gun is on the order of $10^{-5}$ m for a 1-nC bunch. Primarily because of the longitudinal space-charge force associated with high charge density, a relatively long pulse is generated at the dc gun, then an rf buncher following the gun is used to reduce the microbunch length sufficiently to allow injection into one cycle of the accelerating rf, where additional compression can take place. The rf buncher increases the emittance to the $10^{-4}$-m level. RF guns operate with much higher fields at the cathode, which permits generation of short pulses without significant blow up of the longitudinal emittance, which in turn makes an rf buncher unnecessary. A transverse emittance of $1.2 \times 10^{-6}$ m has been demonstrated for an S-band rf gun for a charge of 1-nC bunch and square-shaped bunch length of 9 ps FWHM [1]. In addition, the potential to reduce the transverse emittance in one plane to the $10^{-8}$-m level using an optical transformation [2] is being actively investigated [3].

[*]Work supported by Department of Energy contract DE–AC03–76SF00515.

The proposal here is to adapt an rf gun to the production of polarized electron beams by using a highly p-doped GaAs crystal for the cathode and exciting with circularly polarized laser light of the required wavelength (in the range of 750-850 nm). The plane-wave-transformer (PWT) gun is chosen as a design that best matches the requirements for continuously operating an NEA GaAs crystal in a high rf field for long time periods. These requirements will next be outlined.

## REQUIREMENTS FOR A POLARIZED SOURCE

There is much experience operating GaAs crystals in dc guns biased at about 100 kV [4]. An atomically clean GaAs crystal must be activated with Cs and an oxidizer to produce a negative electron affinity (NEA) surface. This is best done in a UHV chamber isolated from the gun by a gate valve, after which the crystal must be inserted into the gun without breaking vacuum. For an rf gun, the crystal diameter must be kept small, typically about 1-cm, and the holder designed to eliminate rf breakdown. To produce a low emittance beam, the laser spot on the cathode should have a radius of about 1 mm and a pulse length that is $\leq 20°$ rf phase, which for S-band is $\leq 20$ ps. Since this results in extremely high current densities at the cathode, care must be taken to choose a cathode design that is unaffected by the surface charge limit [5]. For high polarization, the active layer of the GaAs crystal will be only about 100 nm thick, which ensures that the low-charge extraction time is only a few picoseconds [6].

Successful dc guns operate with a vacuum of $<10^{-11}$ Torr. The best pressures achieved with presently operating rf guns are on the order of $10^{-10}$ Torr. Some gas species, such as $H_2O$ and $CO_2$, affect the quantum efficiency (QE) of the cathode more than others, such as $H_2$ or CO. The dynamics of contamination and surface damage of the GaAs by gas molecules and ions formed in the rf cavity have yet to be explored experimentally.

HV breakdowns in dc guns have a devastating effect on the QE. One can assume the same will happen with rf breakdowns. Thus the gun rf cavities must be designed and processed to eliminate this phenomenon. The rf processing must be done with a dummy cathode. Based on experience with dc guns, the resulting dark current must be $\ll 1$ nC per $\mu$s of rf. The large dark current observed in the BINP experiment with an activated GaAs cathode in a prototype S-band rf gun was attributed to secondary electron emission [7], which can be high for NEA GaAs. Reducing electron bombardment as well as keeping the emission time short may prove to be crucial.

Magnetic fields at the cathode can potentially decrease the beam polarization. Since the initial spin vector is axial, transverse magnetic fields are the principal concern. Azimuthal rf magnetic fields, $B_\theta(r)$, exist on the cathode during extraction that are of zero magnitude at $r=0$ and increase as $\sim r/R_0 \, E_0 \sqrt{\varepsilon\mu}$ up to $R_0/2$, where $E_0$ is the electric field on the cathode at the time of extraction and $R_0$ is the radius of the cavity. In this proposal these fields are negligible at $r=1$ mm $\approx 0.025 R_0$.

## THE INTEGRATED PWT PHOTOINJECTOR

An example of an S-band integrated PWT photojector is shown in Fig. 1. It consists of a standing wave structure of ½+10+½ cells operating in the π-mode. The disks, which are suspended from 4 water-carrying rods, are mounted inside a large cylindrical tank. The rf power couples first into the annular region of the tank in a TEM-like mode, then couples into the accelerating cells in a TM-like mode on axis—thus the name.

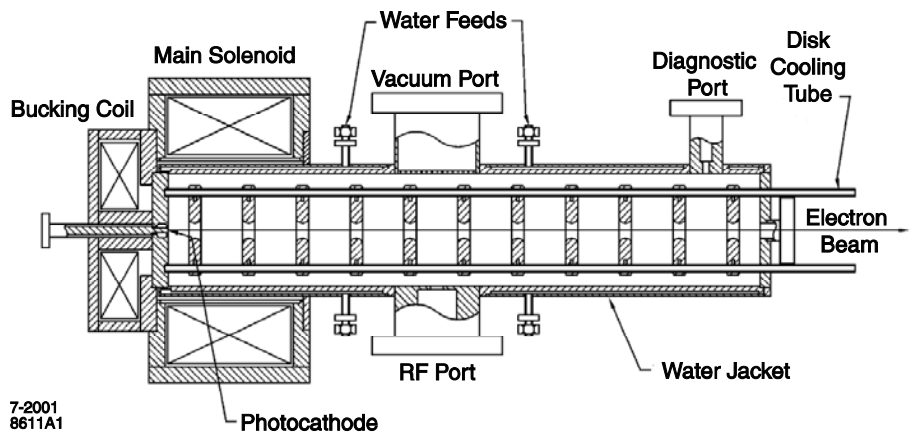

**FIGURE 1**. The integrated PWT photoinjector.

There are several reasons for choosing the PWT design over the 1.6-cell gun plus booster. Parmela simulations indicate that a peak field of 120-140 MV/m at the cathode is required to achieve a transverse emittance of $10^{-6}$ m for a 1-nC bunch, 10 ps pulse. However, the growth in emittance as the peak field is lowered to 60 MV/m is much less for the PWT injector. Thus for collider applications the PWT injector can be operated at lower fields, which generally will reduce dark current and avoids any field emission from the GaAs due to an inversion layer.[1] In addition, the separation of the tank from the disks improves the conductance for vacuum pumping and also allows a wide range of Q values and thus filling times. Long filling times may reduce rf breakdown.

For polarized beam applications, the PWT gun shown in Fig. 1 can be improved as follows. To reduce dark current and rf breakdown, Class 1 OFHC Cu forged using the hot isostatic pressure (HIP) method and single-point diamond machining to a roughness of 0.5 µm or better can be used. The final assembly should undergo a simple rinsing in ultra-pure water [8]. Vacuum pumping can be improved by increasing the tank diameter and coating its inner surface with a thin film getter material such as TiZr or TiZrV [10]. The increased tank diameter will also increase the stored energy, which is desirable for pulse trains. It is possible to increase the water cooling to handle 25 MW peak rf power at 180 Hz. The design parameters for a prototype PWTS-band polarized photoinjector are given in Table 1.

---

[1] When the field into the surface of a p-doped semiconductor is strong enough that the conduction band at the surface is bent closer to the Fermi level than the valence band, then the surface becomes n-type with electrons occupying states in the conduction band near the surface. See ref. [9] for a fuller discussion.

Table 1. S-Band PWT Design Parameters

| Parameter | Value |
|---|---|
| Frequency | 2856 MHz |
| Energy | 20 MeV |
| Charge per bunch | 1 nC |
| **Normalized emittance** | **1 µm** |
| Energy spread | <0.1% |
| Bunch length | 2 ps rms |
| Rep. rate | 5 Hz |
| Peak current | 100 A |
| Linac length | 58 cm |
| Beam radius | <1 mm |
| Peak B field | 1.8 kG |
| **Peak gradient** | **60 MV/m** |
| Peak brightness | $2 \times 10^{14}$ A/(m-rad)$^2$ |

GaAs crystals must be prepared to have an NEA surface in a separate UHV chamber, and then inserted into the gun without venting. Such systems are in routine use for dc-biased guns, but will require some careful development for an rf gun since an rf seal must be made while the crystal itself, which is quite fragile, must not undergo significant stress.

## FUTURE PLANS AND CONCLUSIONS

An SBIR-I has recently been awarded to DULY Research, Inc. to begin the engineering design of an S-band PWT integrated photoinjector with load lock for polarized electron beams. If this is followed by a Phase II award, the photoinjector will be tested at the Gun Test Facility at SLAC. The test will be for QE in a high rf field of an activated GaAs cathode at 20 MeV. The QE during the test will be monitored using a green laser (532 nm) of 2-10 ps pulse length. Polarization and emittance measurements will be performed in subsequent testing.

A PWT integrated rf photoinjector built specifically for polarized electrons has many features that will be important for successful operation of polarized electron beams. Such a gun can be built and tested in the near future.